\documentclass[fleqn,twoside]{article}
\usepackage{espcrc2}

\usepackage{graphicx}
\usepackage[figuresright]{rotating}

\usepackage{epsfig}

\usepackage{bm}

\newcommand{\AmS}{{\protect\the\textfont2
  A\kern-.1667em\lower.5ex\hbox{M}\kern-.125emS}}

\newcommand{\kpc}{{\rm kpc}}
\newcommand{\Mpc}{{\rm Mpc}}

\newcommand{\xiklyaconcordfig}{
    \begin{figure}[htb]
    \epsfxsize=2.9in
    \epsfbox{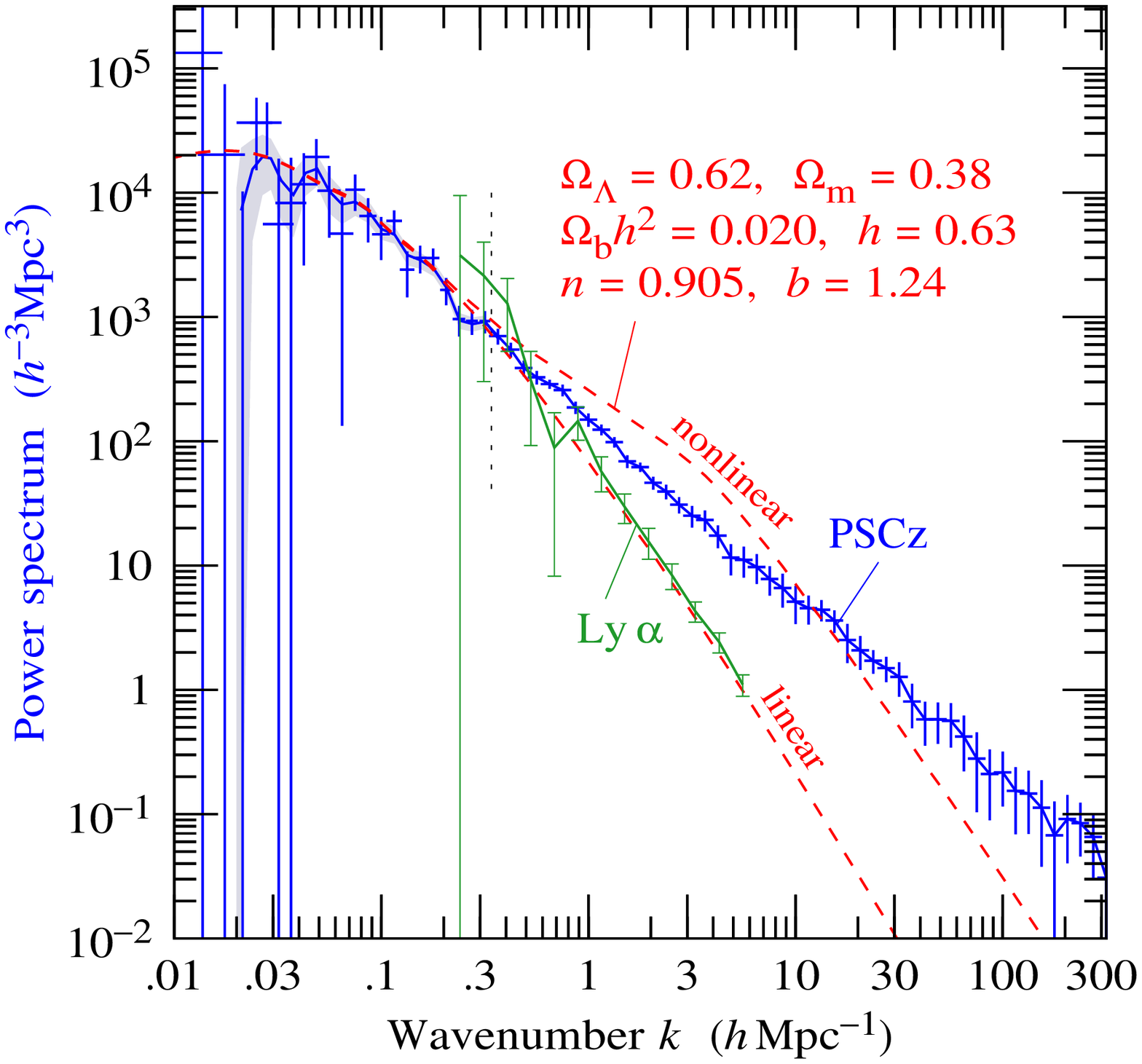}
    \vspace{-8pt}
\caption[1]{\small
The linear matter power spectrum inferred from the Ly$\, \alpha$ forest
\protect\cite{GH02},
compared to the PSCz \protect\cite{HT02} galaxy power spectrum
and to the $\Lambda$CDM concordance model of \protect\cite{TZH01},
both linear and
nonlinearly evolved by the method of \protect\cite{PD}.
    \label{xiklyaconcord}
    }
    \end{figure}
}

\newcommand{\xikcontsfig}{
    \begin{figure*}[htb]
    \begin{minipage}{\textwidth}
    \begin{center}
    \leavevmode
    \epsfxsize=4in	
    \epsfbox{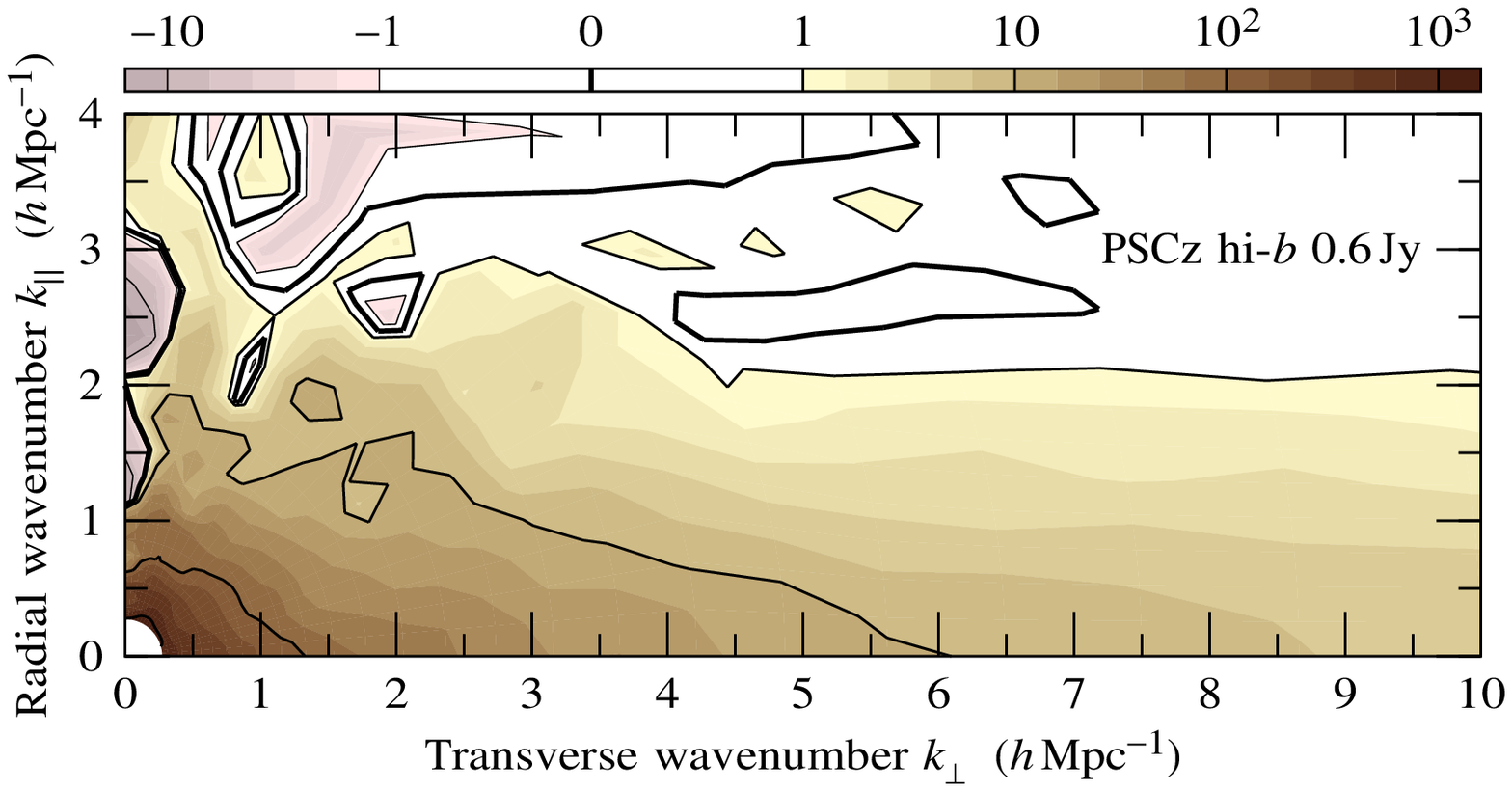}
    \end{center}
    \vspace{-16pt}
    \caption[1]{\small
Contour plot of the redshift space power spectrum
$P^s(k_\perp,k_\parallel)$
of the PSCz 0.6~Jy survey at nonlinear scales
\cite{HT02}.
Power along the transverse (horizontal) axis is unaffected by redshift
distortions, and is therefore equal to the real space power spectrum.
Velocity dispersion suppresses power away from the transverse axis.
Thick, medium, and thin contours represent
zero, positive, and negative values respectively.
    \label{xikconts}
    }
    \end{minipage}
    \end{figure*}
}

\newcommand{\xikblfig}{
    \begin{figure}[htb]
    \epsfxsize=2.9in
    \epsfbox{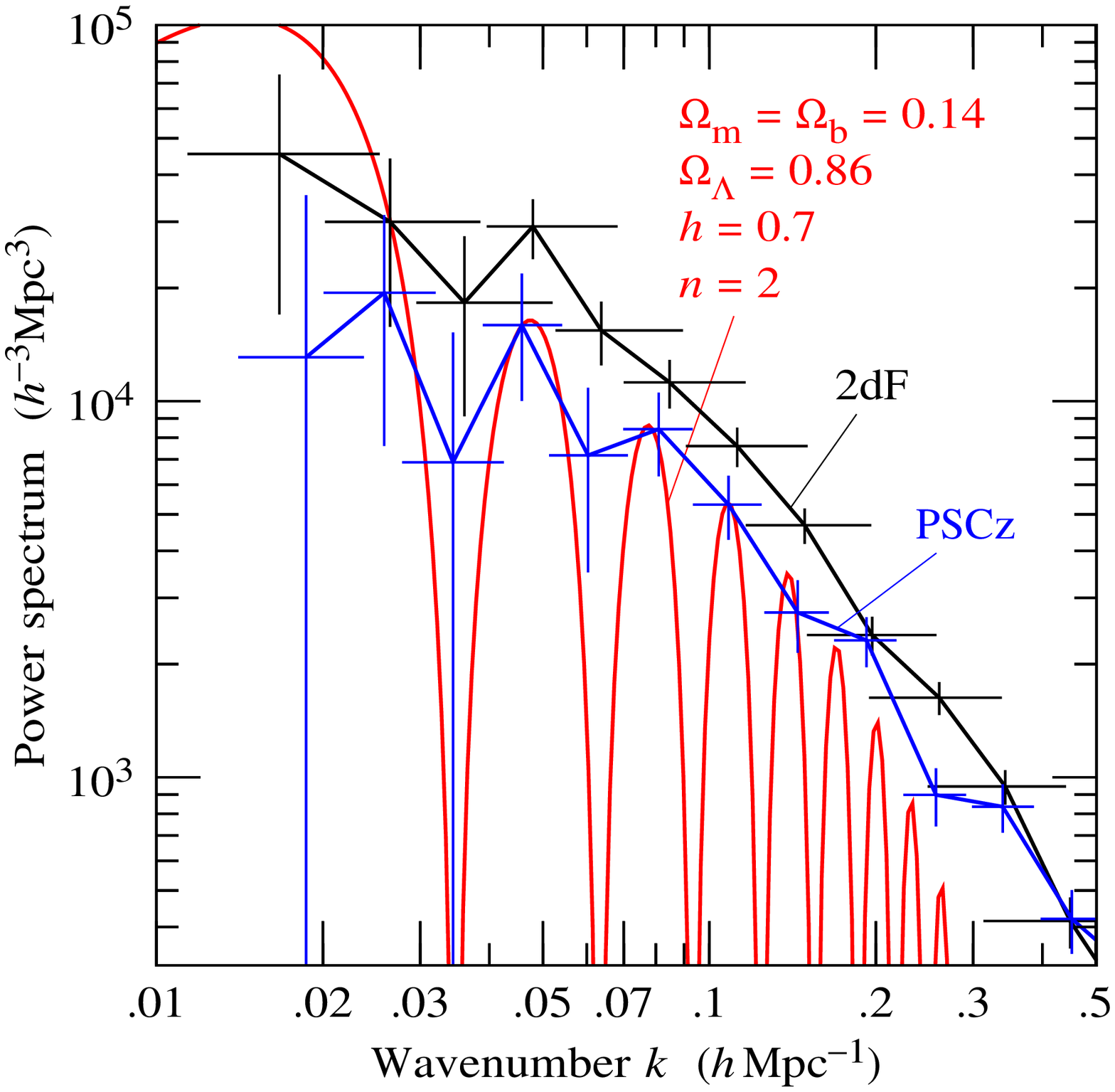}
    \vspace{-8pt}
    \caption[1]{
The 2dF \protect\cite{THX02}
and PSCz \protect\cite{HTP00} galaxy power spectra,
compared to a flat, COBE-normalized
$\Lambda$-baryon model power spectrum
\cite{EH}
with $\Omega_\Lambda = 0.86$ and $\Omega_b = 0.14$,
and no non-baryonic Dark Matter.
    \label{xikbl}
    }
    \end{figure}
}

\title{Cosmology from Large Scale Structure}

\author{Andrew J. S. Hamilton\address[JILA]{JILA and Dept.\ of Astrophysical and Planetary Sci.,
        Box 440, U. Colorado, Boulder, CO 80309, USA}
        \thanks{Andrew.Hamilton@colorado.edu \newline http:$/\!/$casa.colorado.edu/$\sim$ajsh/},
	Nick Gnedin\address[CASA]{CASA and Dept.\ of Astrophysical and Planetary Sci.,
        Box 389, U. Colorado, Boulder, CO 80309, USA}
	\thanks{gnedin@casa.colorado.edu \newline http:$/\!/$casa.colorado.edu/$\sim$gnedin/},
        Max Tegmark\address[PENN]{Dept.\ of Physics, Univ. of Pennsylvania, Philadelphia, PA 19104, USA}
        \thanks{max@physics.upenn.edu \newline http:$/\!/$www.hep.upenn.edu/$\sim$max/},
	and
	Yongzhong Xu\addressmark[PENN]
	\thanks{xuyz@hep.upenn.edu}
}

\begin{document}

\begin{abstract}
We report on
the linear matter power spectrum reconstructed from the Ly$\, \alpha$ forest,
on baryonic wiggles in the galaxy power spectrum,
and on how to measure real space (as opposed to redshift space) power.
\vspace{1pc}
\end{abstract}

\maketitle

\section{INTRODUCTION}

Much of what I presented at the DARK2002 meeting
is contained in \cite{GH02,HT02,THX02,TZH01,Wang01}.
To avoid redundancy,
I will not repeat what is in those papers,
but instead will devote this space to
a subset of issues addressed at the meeting,
namely
the linear matter power spectrum reconstructed from the Ly$\, \alpha$ forest,
baryonic wiggles in the galaxy power spectrum,
and how to measure real space (as opposed to redshift space) power.

\xiklyaconcordfig

\section{POWER FROM THE LYMAN ALPHA FOREST}

In a remarkable paper,
Croft et al.\ \cite{Croft}
reconstructed the linear matter power spectrum
from the measured power spectrum of transmitted flux
in the Lyman alpha forests of a sample of 53 Keck spectra
of quasars at redshifts $z = 2.2$--$4.1$.
The Ly$\, \alpha$ forest offers two advantages over galaxies
as a way to access the linear matter power spectrum.
First,
the scale at which objects are going nonlinear is smaller
at high redshift.
And second,
the Ly$\, \alpha$ forest is less biased,
or at least we think we understand the bias better,
than galaxies.

Some of us were much impressed with Croft's technique,
but at the same time skeptical that it could work so well.
The correction from flux power spectrum to linear matter power
is large and at least somewhat model-dependent,
yet the error bars on the reconstructed matter power spectrum
reported in \cite{Croft}
were tiny at the smallest measurable scales.

Nick Gnedin and I \cite{GH02}
decided to try to reproduce Croft et al.'s \cite{Croft} result,
and Figure~\ref{xiklyaconcord}
shows the result.
We explored a wider range of cosmological parameters,
and astrophysical parameters governing
the relation between dark matter and transmitted flux.
The error bars in Figure~\ref{xiklyaconcord}
on the linear matter power spectrum reconstructed
from the Ly$\, \alpha$ forest
include both statistical and `systematic' errors,
the systematic errors being the envelope of uncertainty
from differently parametered models
that fit the observed power spectrum of transmitted flux.
Although our error bars are bigger than those of \cite{Croft},
we confirm that their procedure seems to work as advertised.
Our conclusions in this regard are more optimistic than those of
Zaldarriaga et al.\ \cite{ZSH}.

For comparison,
Figure~\ref{xiklyaconcord}
also shows the real space power spectrum of PSCz galaxies
measured by \cite{HT02},
and the $\Lambda$CDM concordance model power spectrum
of \cite{TZH01}.
The linear matter power spectrum reconstructed from the Ly$\, \alpha$ forest
has been linearly evolved to the present time using the concordance model.
The agreement between the power spectrum inferred from the Ly$\, \alpha$
forest and the concordance model is quite striking
when you consider that the concordance model was obtained
without the benefit of the Ly$\, \alpha$ data.

\xikblfig

\section{A BARYONIC UNIVERSE?}

One of the hottest goals
(e.g.\ \cite{EHT})
of studies of large scale structure
is to detect baryonic wiggles in the galaxy power spectrum
In \cite{Percival},
the 2 degree Field team reported a tentative detection of baryonic wiggles.
Our own analysis \cite{HT02},
which however is based only on the smaller published 100k data set
\cite{Colless},
finds a similar bump,
but does not rate it as a statistically significant detection of baryons.

Nevertheless the bump at $k \approx 0.05 \, h \, \Mpc^{-1}$
is interestingly large.
Intriguingly, it lines up nicely with a similar bump
in the PSCz power spectrum \cite{HTP00},
as shown in Figure~\ref{xikbl}.
The 2dF and PSCz power spectra in Figure~\ref{xikbl}
are decorrelated \cite{HT00},
meaning that the error bars are uncorrelated with each other,
so it is legitimate to do chi-by-eye.

For fun,
Figure~\ref{xikbl}
compares the 2dF and PSCz power spectra
to a COBE-normalized, flat, adiabatic power spectrum
containing a cosmological constant and baryons,
$\Omega_\Lambda = 0.86$, $\Omega_b = 0.14$,
but no non-baryonic Dark Matter.
A large baryonic component not only strengthens baryonic wiggles,
but also steepens the power spectrum.
To counteract this steepening,
the model power spectrum in Figure~\ref{xikbl}
has a large blue tilt,
with primordial spectral index $n = 2$.

A comparably decent fit can be obtained without a cosmological constant,
but only by dint of increasing the bias factor to the extreme value of
$b \approx 11$.
In other words,
in a pure baryonic universe with no cosmological constant,
Large Scale Structure fluctuations
would predict CMB fluctuations 11 times larger than observed.
This discrepancy is of course an ancient result,
one of the original motivations for introducing the notion
of non-baryonic Dark Matter.

Sadly the $\Lambda$-baryon model fails dismally
compared with CMB observations:
the model predicts a first acoustic peak
5 times larger than observed,
and at harmonic number $l \approx 310$,
way higher than the observed peak at $l = 210$ \cite{M02}.

\xikcontsfig


\section{REAL SPACE POWER}

The PSCz power spectrum shown in Figure~\ref{xiklyaconcord}
is a real space power spectrum, not a redshift space power spectrum.
That is, the effect of pairwise peculiar velocities has been eliminated,
at both linear and nonlinear scales.

Quiz question for you:
at nonlinear scales,
which can you measure more accurately from a galaxy redshift survey,
the (angle-averaged) redshift space power spectrum
or the real space power spectrum?

The answer is obvious, right?
The redshift space power spectrum must be more accurate,
because the observations are in redshift space,
and to get from redshift space to real space
you have to deconvolve the redshift space power spectrum
from the effects of pairwise peculiar velocities.
Presumably that deconvolution is model-dependent,
and liable to the kind of numerical instability
that generally attends deconvolution.

Wrong.

Remarkably enough,
at nonlinear scales
the real space power spectrum can be measured more accurately
than the redshift space power spectrum!
Which is a wonderful thing,
because it is the real space power spectrum that you really want\footnote{
Ok, so what you {\em really} want is a power spectrum
free not only from redshift distortions, but also from bias.
The thorny issue of bias
goes beyond the scope of this commentary.
}.
As an added bonus, the distribution of pairwise peculiar
velocities (or rather its Fourier transform)
emerges as a byproduct of the measurement
(Fig.~2 of \cite{HT02}).

The key point that allows this miracle to happen is that
peculiar velocities displace galaxies
only in the line-of-sight direction in redshift space.
This has the consequence that Fourier modes transverse to the line of sight
are completely unaffected by redshift distortions.
The result is familiar to the executors of angular galaxy surveys,
who know that
the 3D power spectrum
that they extract from an angular survey is a real space power spectrum,
not a redshift space power spectrum.

It follows that
the real space power spectrum
$P(k)$
is equal to
the redshift space power spectrum
$P^s(k_\perp, k_\parallel)$
in the transverse ($k_\parallel = 0$) direction
\begin{equation}
\label{Ps0}
  P(k)
  =
  P^s(k_\perp{=}k, k_\parallel{=}0)
  \ .
\end{equation}
Notice that it is the power spectrum,
not the correlation function,
for which the relation between real space and redshift space quantities
is so simple, equation~(\ref{Ps0}).
If you want to see the mathematics in more detail,
look at \cite{HT02}.

Figure~\ref{xikconts}, from \cite{HT02},
shows a contour plot of the redshift space power spectrum
$P^s(k_\perp,k_\parallel)$
of PSCz,
as a function of wavenumbers $k_\perp$ and $k_\parallel$
respectively transverse and parallel to the line-of-sight.
According to equation~\ref{Ps0},
the real space power spectrum
is equal to the redshift space power spectrum along the transverse axis
in Figure~\ref{xikconts}.

A simple argument demonstrates how it can be that the
real space power spectrum is more accurately measurable than
the redshift space power spectrum.
Take a close pair of galaxies, say $30 \, h^{-1} \kpc$ apart.
Projected on the sky,
the galaxies will appear $30 \, h^{-1} \kpc$ apart or less.
But thanks to their relative peculiar velocities,
the pair will appear typically $3 \, h^{-1} \Mpc$ apart
in redshift space, a factor of 100 larger.
You see that all that wonderful information about close pairs
that is in present in the transverse power spectrum
is horribly mixed into (100 times, for example) larger scales 
in the angle-averaged redshift space power spectrum.



%

\section{CONCLUSIONS}

Cosmologists now have a Standard Model,
a flat $\Lambda$CDM model seasoned with baryons.
At the DARK2002 conference
several speakers reported how data from diverse sources
are pointing at the same concordance model.

Figure~\ref{xiklyaconcord}
illustrates one example of this concordance,
how the linear power spectrum inferred from
the Ly$\, \alpha$ forest
is in full agreement with the concordance model.

First results from the new generation of large redshift surveys
are emerging.
It is probably premature to claim a detection of baryonic wiggles,
but there are tantalizing hints of such,
Figure~\ref{xikbl},
and it should not be long before baryonic wiggles in the galaxy
power spectrum become an observational reality.

This commentary concluded
by bringing attention to the counter-intuitive fact that it is possible
to measure the real space power spectrum of galaxies
more accurately than the redshift space power spectrum.
One hopes that those who analyse redshift surveys will begin
to take advantage of this have-your-cake-and-eat-it truth.

\end{document}